\documentclass[aps,prd,twocolumn,groupedaddress,superscriptaddress, twoside,notitlepage,preprintnumbers,showpacs,
showkeys,byrevtex,floatfix,amsmath,amssymb]
{revtex4-1}
\usepackage{array,amscd,amsmath,amssymb,graphicx,mathtools, multirow,placeins,slashed,verbatim,subfig,tensor,dsfont}
\usepackage[bookmarksnumbered,bookmarksopen=true,bookmarksopenlevel=1,pdfstartview=FitH]{hyperref}
\usepackage[all]{hypcap}
\usepackage{rotating}
 \usepackage{cleveref}
\newcolumntype{M}[1]{>{$}{#1}<{$}}

\newcommand{\sst}[1]{{\scriptscriptstyle #1}}

\def\0{{\sst{(0)}}}
\def\1{{\sst{(1)}}}
\def\2{{\sst{(2)}}}
\def\3{{\sst{(3)}}}
\def\4{{\sst{(4)}}}
\def\5{{\sst{(5)}}}
\def\6{{\sst{(6)}}}
\def\7{{\sst{(7)}}}

\newcommand{\be}{\begin{equation}}
\newcommand{\ee}{\end{equation}}
\def\ba{\begin{array}}
\def\ea{\end{array}}

\newcommand\half{\tfrac{1}{2}}

\newcommand{\bea}{\begin{eqnarray}}
\newcommand{\eea}{\end{eqnarray}}

\DeclareMathOperator{\tr}{tr}
\DeclareMathOperator{\Tr}{Tr} 

\newcommand{\F}{\mathds{F}}

\newcommand{\Q}{\mathcal{Q}}

\newcommand{\N}{\mathcal{N}}
\newcommand{\hi}{H_\infty}

\newcommand{\FTS}{\mathfrak{F}}
\newcommand{\J}{\mathfrak{J}}

\begin{document}

\title{Freudenthal duality and conformal isometries of extremal black holes}

\author{L. Borsten}
\email[]{leron@stp.dias.ie}
\affiliation{School of Theoretical Physics, Dublin Institute for Advanced Studies,
10 Burlington Road, Dublin 4, Ireland}
\author{M. J. Duff}
\email[]{m.duff@imperial.ac.uk}
\affiliation{Theoretical Physics, Blackett Laboratory, Imperial College London,
London SW7 2AZ, United Kingdom}
\affiliation{Mathematical Institute, University of Oxford, Andrew Wiles Building, Woodstock Road, Radcliffe Observatory Quarter,
Oxford, OX2 6GG, United Kingdom}
\affiliation{Institute for Quantum Science and Engineering and Hagler Institute for Advanced Study, Texas A\&M University, College Station, TX, 77840, USA}
\author{A. Marrani}
\email[]{jazzphyzz@gmail.com}
\affiliation{Museo Storico della Fisica e Centro Studi e Ricerche ``Enrico Fermi'',
Via Panisperna 89A, I-00184, Roma, Italy}
\affiliation{Dipartimento di Fisica e Astronomia ``Galileo Galilei'', Universit\`a di Padova, and INFN, sezione di
Padova,
Via Marzolo 8, I-35131 Padova, Italy}
\date{\today}

\begin{abstract}

We present  a conformal isometry for static extremal black hole solutions in all four-dimensional Einstein-Maxwell-scalar theories with electromagnetic duality groups `of type $E_7$'. This includes, but is not limited to, all supergravity theories with $\N>2$ supersymmetry and all $\N=2$ supergravity theories with symmetric scalar manifolds. The conformal isometry is valid for  arbitrary electromagnetic charge configurations and relies crucially on the notion of Freudenthal duality. 

\end{abstract}


\preprint{DFPD/2018/TH/04, DIAS-STP-18-18, Imperial-TP-2018-MJD-02}

\maketitle

\section{Introduction} 

The stability of classical solutions to Einstein's equations remains an open question of much importance. Recent years have seen  significant progress for black hole solutions. In particular, the  Schwarzschild solution has been shown to be stable against linear perturbations \cite{Dafermos:2008en, Dafermos:2010hd, Dafermos:2016uzj}. Similarly, for the non-extremal Reissner-Nordstr\"om and Kerr black holes, an analog of the horizon redshift effect can be used to establish the decay of linear scalar perturbations \cite{Dafermos:2008en, Dafermos:2010hd, Dafermos:2014cua}. In the extremal case, however, the surface gravity, and so the  horizon redshift effect, vanishes. Indeed,   Aretakis  demonstrated that extremal Reissner-Nordstr\"om solutions have an unavoidable instability on $\mathcal{H}^+$, the future horizon \cite{Aretakis:2011ha, Aretakis:2011hc,Aretakis:2012ei, Aretakis:2018dzy}. There exists an infinite set of conserved quantities, the Aretakis charges, on  $\mathcal{H}^+$, which are linear in  the scalar perturbation $\psi$.  The Aretakis charges imply that $\partial_r \psi$ does not decay on $\mathcal{H}^+$. However, $\partial_r \psi$ \emph{does} decay off $\mathcal{H}^+$ and, so, $\partial_{r}^{2} \psi$ must diverge on $\mathcal{H}^+$, seeding an instability (see also \cite{Akhmedov:2018okp}).

Supersymmetric extremal  black holes solutions in $D=4$ spacetime dimensions have played, and continue to play, a central role in  string/M-theory. Static  solutions of this kind are structurally  close to  Reissner-Nordstr\"om  and, so, their classically stability would appear to be subject to the arguments of Aretakis. Supersymmetry does not save you. Indeed, even for the maximally supersymmetric  Minkowski vacuum, its stability  is \emph{not} actually implied by the positive energy theorem \cite{Christodoulou:1993uv}.  For  $D=4$ single-centre extremal  black hole   with a Bertotti-Robinson $\text{AdS}_2\times S^2$ near horizon geometry,  which covers all examples known to us, including the supersymmetric solutions, there is an  instability \cite{Lucietti:2012sf}. These results have by now been generalised to a very large class of extremal solutions  and sources of linear perturbations \cite{Lucietti:2012sf, Lucietti:2012xr}.

Returning to the Reissner-Nordstr\"om case, an intuition for the  presence of the instability  is provided  \cite{Lucietti:2012xr} by the Couch-Torrence conformal inversion \cite{couch1984conformal}, which exchanges  $\mathcal{H}^+$ with future null infinity, $\mathcal{I}^+$.  Under such a conformal inversion  the Newman-Penrose charges \cite{Newman:1968uj} at infinity  are sent to Aretakis charges on the horizon, thus suggesting the presence of an instability \cite{Lucietti:2012xr}. For recent developments, see \cite{Cvetic:2018gss}. A  Couch-Torrence-type  conformal inversion has been shown to exist for supersymmetric black hole solutions in very special cases \cite{Godazgar:2017igz, Chow:2017hqe}.   
  In this brief note, we generalise the Couch-Torrence-type  conformal inversion to arbitrary  extremal, static,  black hole solutions in all Einstein-Maxwell-scalar  theories with electromagnetic duality group $G$ of type $E_7$. This includes \emph{all} $\mathcal{N}$-extended supergravities with $\mathcal{N}>2$ supersymmetries, as well as all $\mathcal{N}=2$ theories for which the scalar fields parametrise a symmetric space. Interestingly, the definition of the conformal isometry depends on the notion of \emph{Freudenthal duality}, a non-linear anti-involution of the electromagnetic charges carried by the black hole solution that leaves the Bekenstein-Hawking entropy invariant \cite{Borsten:2009zy, Ferrara:2011gv, Borsten:2012pd}.

\section{Einstein-Maxwell-scalar theories and groups of type $E_7$}

We consider theories of Einstein gravity coupled to Abelian gauge potentials and scalars defined  by $G$ a group `of type $E_7$' (introduced below). In particular, the Abelian field strengths  and their duals transform in a symplectic representation of $G$, while the scalars parametrise the coset $G/K$, where $K$ is the maximal compact subgroup of $G$. This class of theories includes  (the bosonic sector of) all $D=4$ supergravities with $\N>2$ supersymmetries, as well as all $\N=2$ supergravities coupled to vector multiplets with scalars parametrising a symmetric space. Note, however, supersymmetry is not necessary and there are many examples that do not admit a supersymmetric completion, at least in Lorentzian space-time signature. 

\subsection{Groups of type $E_7$}

Groups of type $E_7$ can be characterised by Freudenthal triple systems (FTS). An FTS may be axiomatically defined \cite{Brown:1969} as a finite
dimensional vector space $\FTS$ over a field $\F$ (not of
characteristic 2 or 3), such that:
\begin{enumerate}
\item  $\FTS$ possesses a non-degenerate antisymmetric bilinear form $\{x, y\}.$
\item $\FTS$ possesses a symmetric four-linear form $q(x,y,z,w)$ which is not identically zero.
\item If the ternary product $T(x,y,z)$ is defined on $\FTS$ by $\{T(x,y,z), w\}=q(x, y, z, w)$, then
\be
3\{T(x, x, y), T(y,y,y)\}=\{x, y\}q(x, y, y, y).
\ee
\end{enumerate}
For notational convenience, let us introduce 
\begin{subequations}\label{delta}
\begin{align} 
2\Delta(x, y, z, w)&\equiv q(x, y, z, w),  \\
\Delta(x)& \equiv\Delta(x, x, x, x),\\
T(x)& \equiv T(x, x, x).
\end{align}
\end{subequations}

For a given $\FTS$ the \emph{automorphism} group $G$  is defined  as the set of invertible  $\F$-linear transformations, $\sigma$, preserving the quartic and quadratic forms:
\be
\{\sigma x, \sigma y\}=\{x, y\},\qquad\Delta(\sigma x)=\Delta(x)\label{eq:brownfts}.
\ee
The prototypical example is given by  $G\cong E_7$, in which case $\FTS$ is the fundamental 56-dimensional $E_7$ representation and  the antisymmetric bilinear form and symmetric four-linear form are the unique symplectic quadratic and totally symmetric invariant, respectively. 

We shall also make use of the notion of the Freudenthal dual \cite{Borsten:2009zy, Ferrara:2011gv, Borsten:2012pd},
\be\label{fdual}
\tilde{x} =\varepsilon_x \frac{T(x)}{\sqrt{\Delta(x)}}, \quad  \varepsilon_x:=\text{sgn}[\Delta(x)]
\ee
which is defined for all $x\in\FTS$ such that $\Delta(x)\not=0$ and satisfies 
\be
\Delta(\tilde{x})=\Delta(x), \quad \{\tilde{x}, x\} =2\sqrt{|\Delta(x)|}, \quad \tilde{\tilde{x}}=-x. 
\ee
Note, here we take $\FTS$ over a field,  whereas in \cite{Borsten:2009zy}  an \emph{integral} FTS was considered, which restricts the space on which \eqref{fdual} in defined.   Interestingly, \eqref{fdual} is directly related to  (generalised) Hitchin functionals for groups of type $E_7$ \cite{Levay:2012tg, Borsten:2012pd}, which also appear in the context of generalised geometry and marginal deformations of $\text{AdS}_4$ solutions. See  \cite{Ashmore:2018npi} and references therein.

The two-dimensional subspace  spanned by an F-dual pair $x, \tilde{x}$ is non-degenerate and closed under the triple-product and defines a sub-FTS, which we denote by $\FTS_x\subset\FTS$ and refer to as the F-plane at $x$. Note, for  $y\in \FTS_x$ given by $y=a x + b \tilde{x}$, $a, b \in \F$ we have \footnote{Formula \eqref{abdual} generalizes the results in Sec. 6 of \cite{Fernandez-Melgarejo:2013ksa}; a detailed treatment will be given elsewhere \cite{Borsten:2018xxx}.}
\be\label{abdual}
\Delta(y)=(a^2+\varepsilon b^2)^2 \Delta(x). 
\ee

For $x$, $\Delta(x)\not=0$ define the linear operator 
\be
L_x : \FTS_{x}^{\perp} \longrightarrow \FTS_{x}^{\perp}; \quad L_{x}(u):= T(x,x,u),
\ee
where $\FTS_{x}^{\perp}$ is the orthogonal complement of $\FTS_{x}$ with respect to the antisymmetric bilinear form. Note,  $L_{x}^{2}=-\Delta(x) \text{Id}$ and $\{L_{x}(u), v\}+\{u, L_{x}(v)\}=0$ \cite{Brown:1969}. 

If there is an element $x \in \FTS$ such that $-\Delta(x)$ is a non-zero square the FTS is said to be \emph{reduced} \cite{Brown:1969}. For reduced FTS it can be assumed that   there is an element $f$ such that $\Delta(f)=-1$ and   \cite{Brown:1969},
\be
\FTS = \F f_+\oplus \F f_- \oplus \J_+\oplus\J_-
\ee
where $\J_\pm$ is the $\pm$-eigenspace of $L_f$ forming a cubic Jordan algebra (see \cref{J})  and 
\be
f_\pm =\frac{1}{2}\left[f\mp T(f)\right].
\ee
We shall write elements in such a basis as 
\be
x=(\alpha, \beta, A, B),
\ee
 where $\alpha, \beta \in \F$ and $A, B\in \J_+, \J_-$, respectively. Then 
 \begin{subequations}
    \begin{equation}\label{eq:bilinearform}
        \{x, y\}:=\alpha\delta-\beta\gamma+\Tr(A,D)-\Tr(B,C),  
                \end{equation}
                for $x=(\alpha, \beta, A, B)$ and $y=(\gamma, \delta, C, D)$ and 
    \begin{equation}\label{eq:quarticnorm}
    \Delta (x)=-\kappa(x)^2-4[\alpha N(A)+\beta N(B)-\Tr(A^\sharp, B^\sharp)]
    \end{equation}
\begin{equation}\label{eq:Tofx}
T(x)=\left(\begin{array}{c}-\alpha\kappa(x)-N(B)\\
\beta\kappa(x)+N(A)\\
-(\beta B^\sharp-B\times A^\sharp)+\kappa(x)A\\
(\alpha A^\sharp-A\times B^\sharp)-\kappa(x)B\\
\end{array}\right)^T,
\end{equation}
\end{subequations} 
where $\kappa(x):=\alpha\beta-\Tr(A,B)$. In the context of Einstein-Maxwell-scalar theories of type $E_7$, as introduced below, this decomposition implies  the existence of a $D=5$ Einstein-Maxwell-scalar theory, with global symmetry given by the reduced structure group of the cubic Jordan algebra $\J$, that yields the corresponding $D=4$ theory upon dimensional reduction on a circle.

\subsection{Einstein-Maxwell-scalar theories of type $E_7$}

By Einstein-Maxwell-scalar theories of type $E_7$ we mean that the electromagnetic duality group is the automorphism group $G$ of some $\FTS$, the Abelian field strengths and their duals take values in $\Lambda^2(M)\otimes\FTS$ and that the scalars parametrise the coset $G/K$. This is the case for all  $\mathcal{N}$-extended supergravities with $\mathcal{N}>2$ supersymmetries, as well as all $\mathcal{N}=2$ theories for which the scalar fields parametrise a symmetric space. Note,  however, for $\N=3$ supergravity (coupled to an arbitrary number of vector multiplets), as well as the minimally coupled $\N=2$ supergravities, the corresponding $\FTS$ are not reduced and their quartic invariant is  degenerate in the sense that it is the  square of a quadratic invariant. The `degeneration' of groups of type $E_7$ is discussed in \cite{Ferrara:2012qp}, where in Sec.~10 it is shown that in these cases Freudenthal duality is nothing but a particular anti-involutive symplectic transformation belonging to $G$. 

The two-derivative Einstein-Maxwell-scalar Lagrangian is uniquely determined by the choice of $G$, although $G$ is only a symmetry of the equations of motion due to electromagnetic duality. However, there exists  a manifestly  $G$-invariant and covariant Lagrangian if one is willing to  accept  a twisted-self-duality constraint that must be imposed in addition to the Lagrangian \cite{Cremmer:1997ct, Borsten:2012pd}. This formalism makes the notation compact and we adopt it here. Let us define the  ``doubled'' Abelian gauge potentials $ \mathcal{A}=(A, B)^T$ transforming as a symplectic vector  of $G$, such that
  \be
 \mathcal{F}  =  d \begin{pmatrix} A \\ B \end{pmatrix},
 \ee
 and introduce the manifestly  $G$-invariant Lagrangian,
  \be\label{doubleL}
 \mathcal{L}= R\star 1 +\frac{1}{4}\tr \left(\star d \mathcal{M}^{-1}\wedge d\mathcal{M}\right) -\frac{1}{4}  \star  \mathcal{F}  \wedge  \mathcal{M}   \mathcal{F},
 \ee
 with constraint \cite{Cremmer:1979up},
 \be\label{constraint}
 \mathcal{F}=\star \Omega \mathcal{M}    \mathcal{F},\qquad \Omega = \begin{pmatrix} 0&\mathds{1}\\ -\mathds{1}&0 \end{pmatrix},
 \ee
 where $\mathcal{M}$ is the scalar coset representative and, in a suitable basis, $\{ \mathcal{F},  \mathcal{G}\}= \mathcal{F}^T\Omega \mathcal{G}$.
 The doubled Lagrangian \eqref{doubleL}, where the potential $\mathcal{A}$ is treated as the independent variable, together with  the constraint \eqref{constraint} is on-shell equivalent to the standard  Lagrangian  \cite{Cremmer:1997ct}.
 
\section{Black hole solutions}

For Einstein-Maxwell-scalar theories with electromagnetic duality group $G$ of type $E_7$, the most general  extremal, asymptotically flat, spherically symmetric,
static, dyonic black hole metric is given by (cf.~for example \cite{Bellucci:2008sv} and the references therein)
\begin{equation}
ds^{2}=-e^{2U}dt^{2}+e^{-2U}(dr^{2}+r^{2}d\Omega _{2}),  \label{metric}
\end{equation}%
where $U=U(H\left( r\right))$ and
\be
e^{-2U}=\sqrt{\Delta \left( H \right) }, \quad 
H\left( r\right)=H_{\infty }-\frac{\mathcal{Q}}{r}.
\ee%
Here $H_{\infty }$ and $\mathcal{Q}$ are symplectic vectors of $G$.
Consistency with the scalar equations of motion implies \footnote{Here, we are actually considering those extremal black hole solutions supported by electro-magnetic charges $\mathcal{Q}$ belonging to a particular orbit of the non-transitive action of $G$ over $\mathfrak{F}$. When embedded into supergravity, these solutions are supersymmetric, and of 1/$\mathcal{N}$-BPS type, where $\mathcal{N}$ is the number of supercharges. Extensions to other orbits of the action of $G$ on $\mathfrak{F}$, as well as to other supergravity theories, will be considered elsewhere."} 
\begin{subequations}\label{con}
\begin{align} \label{eom1}
\Delta \left( H\right) &>0, \quad \forall r\in \lbrack
0,\infty ),  \\ \label{con1}
\Delta \left( H_{\infty }\right)
&=1, \\ \label{eom2}
\left\{ H_{\infty },\mathcal{Q}\right\} &=0.
\end{align}
\end{subequations}
The Abelian two-form fields strengths are given by 
\begin{equation}
\mathcal{F}  = \frac{e^{2U}}{r^{2}}\Omega \mathcal{%
M}\mathcal{Q}%
dt\wedge dr+\mathcal{Q}\sin \theta d\theta \wedge d\varphi. \label{vectors}
\end{equation}%
such that%
\begin{equation}
\frac{1}{4\pi }\int_{S_{\infty }^{2}}\mathcal{F}=\mathcal{Q}.\label{flux}
\end{equation}
For $\mathcal{N}=2$ supergravity theories the scalars are given by 
 \begin{equation}\label{scalars}
z^{a}\left( H(r)\right) =\frac{H^{a} -i{\tilde{H}}^{a}}{H^{0} -i{\tilde{H}}^{0}}.  
\end{equation}
Since the scalars are specified by $\Q$ and $\hi$ alone, it is  convenient to introduce the notation,
\be
z^a = z^a(r, \Q, \hi).
\ee
Note, the horizon and asymptotic values of the scalars are respectively given by 
\be
\lim_{r\rightarrow \infty} z^a = \frac{\hi^{a} -i{\tilde{H}_\infty}^{a}}{{H}_{\infty}^{0} -i\tilde{H}_{\infty}^{0}},\qquad \lim_{r\rightarrow 0} z^a = \frac{\Q^{a} -i{\tilde{\Q}}^{a}}{{\Q}^{0} -i\tilde{\Q}^{0}}.
\ee

\section{Conformal isometries}

Consider the inversion 
\be\label{inv}
(t, r, \theta, \phi)\mapsto (t, \tilde{r}=\frac{\sqrt{\Delta}}{r}, \theta, \phi),
\ee
where $\Delta\equiv \Delta(\mathcal{Q})$, which implies  
\be\label{tranhom}
H(r) = -\frac{\tilde{r}}{\Delta^{\frac{1}{4}}}\left(\frac{\mathcal{Q}}{\Delta^{\frac{1}{4}}}-\frac{H_\infty\Delta^{\frac{1}{4}}}{\tilde{r} }\right).
\ee
In the following we consider two circumstances under which this inversion yields  a conformal isometry for generic charge configurations $\Q$.

\subsection{Inversion with a charge transformation}
In the first example, we consider the inversion together with a transformation of the charges given by
\be\label{HQmap}
\begin{pmatrix} H_\infty \\ \mathcal{Q} \end{pmatrix}\mapsto
\begin{pmatrix} 0 & \Delta^{-\frac{1}{4}}\\  \Delta^{\frac{1}{4}} &0 \end{pmatrix} \begin{pmatrix} H_\infty \\ \mathcal{Q} \end{pmatrix}.
\ee
Applying  \eqref{HQmap} to \eqref{tranhom} we obtain    
\be\label{Htrans}
H(r) = - \frac{\tilde{r}}{\Delta^{\frac{1}{4}}} H(\tilde{r}) 
\ee
and so
\be
e^{-2U(r)}=\frac{\tilde{r}^2}{\sqrt{\Delta}} e^{-2U(\tilde{r})}.
\ee
Consequently the inversion \eqref{inv} together with the charge map \eqref{HQmap} yields a conformal inversion,
\be
ds^2 = \frac{\sqrt{\Delta}}{\tilde{r}}d\tilde{s}^2,
\ee
where
\be
d\tilde{s}^2=-e^{2U(\tilde{r})}d\tilde{t}^{2}+e^{-2U(\tilde{r})}(d\tilde{r}^{2}+\tilde{r}^{2}d\Omega _{2}).
\ee

Under \eqref{inv} and  \eqref{HQmap} the scalars \eqref{scalars} are invariant,
\be\label{ztrans}
z(r, \Q, \hi)\mapsto z(\tilde{r}, \Q, \hi),
\ee 
since the conformal factor in \eqref{Htrans}  cancels in the ratio \eqref{scalars}. 
On the other hand, the field strengths and their duals undergo a non-trivial transformation 
\[
\mathcal{F}\mapsto - \Delta^{\frac{1}{4}}\left(\frac{e^{2U(\tilde{r})}\Omega \mathcal{M}  \hi }{\tilde{r}^{2}}%
dt\wedge dr-  \hi d\cos \theta \wedge d\varphi\right),
\]
where the scalar coset representative $\mathcal{M}=\mathcal{
M}(z, \bar{z})$ is invariant due to \eqref{ztrans}.

Note, for the special case of $H_\infty \Delta^{\frac{1}{4}}=\Q$ the charge map is an identity operation and so the inversion \eqref{inv} alone is a conformal isometry. This choice corresponds to the doubly-extremal solutions for which the scalars are constant,
 \begin{equation}\label{DCscalars}
z^{a}(r)=\frac{\Q^{a} -i{\tilde{\Q}}^{a}}{\Q^{0} -i{\tilde{\Q}}^{0}}. 
\end{equation}
which also implies 
\be
\mathcal{F}(t, r) \mapsto \mathcal{F}(-{t}, \tilde{r}).
\ee

For both general $(H_\infty, \Q)$ and the special case of $H_\infty \Delta^{\frac{1}{4}}=\Q$, this set-up is not entirely satisfactory. In the first instance, the inversion must be accompanied by a transformation of the charges that is not in the duality group $G$, so is not simply a diffeomorphism. For the latter, the charge transformation is not required, but setting $H_\infty \Delta^{\frac{1}{4}}=\Q$ reduces the solution to the special case of constant scalars, which somewhat trivialises the observation. 

\subsection{Inversion on the F-plane}

One can avoid the need to transform the charges by considering $\hi$ in the orthogonal complement of the F-plane defined by $\Q$,  
\begin{subequations}\label{fplane}
\begin{align} \label{fp1}
\{ H_\infty, \tilde{\mathcal{Q}}\}=0  \\ \label{fp2}
\{ \tilde{H}_\infty, {\mathcal{Q}}\}=0 
\end{align}
\end{subequations}
In this case 
\be
\begin{split}
\Delta(H(r)) & = \frac{\tilde{r}^4}{\Delta}\left[1-\frac{2}{\tilde{r}}\{ \tilde{\mathcal{Q}}, H_\infty\} \right.\\
&\phantom{= \frac{\tilde{r}^4}{\Delta}\left[1\right.}+ \frac{3}{\tilde{r}^2}\{ L_\mathcal{Q}(H_\infty), H_\infty\} \\
&\phantom{= \frac{\tilde{r}^4}{\Delta}\left[1\right.} \left.-\frac{2 \Delta^\frac{1}{2}}{\tilde{r}^3}\{ {\mathcal{Q}}, \tilde{H}_\infty\} +  \frac{\Delta}{\tilde{r}^4}\right]\\
&= \frac{\tilde{r}^4}{\Delta} \Delta(H(\tilde{r}))
\end{split}
\ee
without further intervention and, hence, 
\be
ds^2 = \frac{\sqrt{\Delta}}{\tilde{r}}d\tilde{s}^2.
\ee

In this case the scalars transform non-trivially
\be\label{ztrans2}
z(r, \Q, \hi)\mapsto z(\tilde{r}, \Delta^\frac{1}{4}\hi, \Delta^{-\frac{1}{4}}\Q)\equiv z',
\ee 
and, therefore, so do the field strengths through their dependence on $\mathcal{M}(z, \bar{z})\mapsto \mathcal{M}(z', \bar{z}')$, which implies
\[
\mathcal{F}\mapsto  -\frac{e^{2U(\tilde{r})}}{\tilde{r}^{2}}\Omega
\mathcal{M}\left(z^{\prime},\bar{z}^{\prime}\right) \mathcal{Q}%
dt\wedge d\tilde{r}+\mathcal{Q}\sin \theta d\theta \wedge d\psi.
\]

This construction provides another class of Couch-Torrent-like conformal inversions,  complementing and generalising the case  presented in \cite{Godazgar:2017igz}. It is valid for generic charge configurations (with  asymptotic scalars related through \eqref{fplane}) supporting $\Delta>0$  extremal black hole solutions in all $D=4$ Einstein-Maxwell-scalar theories (not necessarily supersymmetric) with electromagnetic duality group of type $E_7$. The  conformal inversion exchanges the   Newman-Penrose and Aretakis charges,  providing a heuristic argument for the instability \cite{Lucietti:2012xr} in all such theories. There are a number of possible further developments. Firstly, the extension of the above treatment beyond electromagnetic duality groups of type $E_7$ can be considered, even if just a few explicit running scalar flows are known, since the notion of Freudenthal duality can be suitably generalised \cite{Ferrara:2011gv}.
Similarly, the extension to asymptotically non-flat extremal black holes can be considered, for example, in the context of attractors in gauged supergravity theories. See, in particular, \cite{Klemm:2017xxk} and the references therein.  Given the existence of an analogous `Jordan dual' in $D=5$ \cite{Borsten:2009zy}, it may be possible to generalise the conformal map to $D=5$. For a conformal isometry of  black strings in $D>4$  and related recent developments, see also \cite{Chow:2017hqe} and references therein.
Finally, one might also consider generalizations to multi-centered extremal black holes. For example, see \cite{Ferrara:2006yb, Bossard:2011kz} and the references therein.

\appendix

\section{Cubic Jordan Algebras}\label{J}
A Jordan algebra $\mathfrak{J}$ is vector space defined over a
ground field $\mathds{F}$ (not of characteristic 2) equipped with a bilinear product
satisfying \cite{Jordan:1933a, Jordan:1933b, Jordan:1933vh}
\begin{equation}\label{eq:Jid}
A\circ B =B\circ A,\quad A^2\circ (A\circ B)=A\circ (A^2\circ B),
\end{equation}
$\forall\ A, B \in \mathfrak{J}$.
A \emph{cubic norm} is a homogeneous map of degree three
\begin{equation}\label{eq:cubicnorm}
N:V\to \mathds{F}, \quad\text{s.t.} \quad N(\alpha A)=\alpha^3N(A), 
\end{equation}
$\forall \alpha \in \mathds{F}, A\in V$, such that its linearization is trilinear.
Let $V$ be a vector space equipped with a cubic norm. If $V$ further
contains a base point $N(c)=1$  one may define the following
four maps:
\begin{subequations}\label{eq:cubicdefs}
\begin{enumerate}
\item The trace,
    \begin{equation}
    \Tr(A)=3N(c, c, A),
    \end{equation}
\item A quadratic map,
    \begin{equation}
    S(A)=3N(A, A, c),
    \end{equation}
\item A bilinear map,
    \begin{equation}
    S(A, B)=6N(A, B, c),
    \end{equation}
\item A trace bilinear form,
    \begin{equation}\label{eq:tracebilinearform}
    \Tr(A, B)=\Tr(A)\Tr(B)-S(A, B).
    \end{equation}
\end{enumerate}
\end{subequations}
A cubic Jordan algebra $\mathfrak{J}$ with multiplicative identity $\mathds{1}=c$ may be derived from any such vector space if $N$ is \emph{cubic}:
\begin{enumerate}
\item The trace bilinear form \eqref{eq:tracebilinearform} is non-degenerate.
\item The quadratic adjoint map, $\sharp\colon\mathfrak{J}\to\mathfrak{J}$, uniquely defined by $\Tr(A^\sharp, B) = 3N(A, A, B)$, satisfies
    \begin{equation}\label{eq:Jcubic}
    (A^{\sharp})^\sharp=N(A)A, \qquad \forall A\in \mathfrak{J}.
    \end{equation}
\end{enumerate}
The Jordan product is given by
\begin{equation}\label{eq:J3prod}
A\circ B := \half\big(A\times B+\Tr(A)B+\Tr(B)A-S(A,
B)\mathds{1}\big),
\end{equation}
where,
\begin{equation}\label{eq:FreuProduct}
A\times B := (A+B)^\sharp-A^\sharp-B^\sharp.
\end{equation}

\acknowledgments 

We are grateful   to Philip Candelas for hospitality at the Mathematical Institute, University of Oxford.  MJD is also grateful to Marlan Scully for his hospitality in the Institute for Quantum Science and Engineering, Texas A\&M University. He acknowledges the Leverhulme Trust for an Emeritus Fellowship and the Hagler Institute for Advanced Study at Texas A\&M for a Faculty Fellowship. 
 The work of MJD is supported  in part by the STFC under rolling grant ST/P000762/1.  The work of LB is supported by a Schr\"odinger Fellowship.


%

\end{document}